\documentclass[article,twocolumn]{revtex4}

\usepackage{graphicx}
\usepackage{amsmath}
\usepackage{amssymb}
\usepackage{color}
\usepackage{bm}

\begin{document}

\title{Inelastic quantum transport: the self-consistent Born approximation \\and correlated electron-ion dynamics}
\author{Eunan J. McEniry}
\affiliation{Atomistic Simulation Centre, School of Mathematics and Physics, Queen's University Belfast, Belfast BT7 1NN, UK}
\email{e.mceniry@qub.ac.uk}
\author{Thomas Frederiksen}
\affiliation{Donostia International Physics Center (DIPC), Manuel de Landizabal Pasealekua, E-20018 Donostia, Spain}
\affiliation{CIC nanoGUNE Consolider, Mikeletegi Pasealekua 56, E-20009 Donostia, Spain}
\author{Tchavdar N. Todorov}
\author{Daniel Dundas}

\affiliation{Atomistic Simulation Centre, School of Mathematics and Physics, Queen's University Belfast, Belfast BT7 1NN, UK}
\author{Andrew P. Horsfield}
\affiliation{Department of Materials, Imperial College London, South Kensington Campus, London SW7 2AZ, UK}

\date{\today}

\newcommand \diff {{\rm{d}}}
\newcommand \Tr {{\rm {Tr}}}
\newcommand \bfr {{\bf{r}}}
\newcommand \bfp {{\bf{p}}}
\newcommand \bfnabla {{\bm{\nabla}}}
\newcommand \eye {{\rm i}}
\renewcommand \Re {{\rm Re\ }}
\renewcommand \Im {{\rm Im\ }}

\begin{abstract}
A dynamical method for inelastic transport simulations in nanostructures is compared with a steady-state method based on non-equilibrium Green's functions. A simplified form of the dynamical method produces, in the steady state in the weak-coupling limit, effective self-energies analogous to those in the Born Approximation due to electron-phonon coupling. The two methods are then compared numerically on a resonant system consisting of a linear trimer weakly embedded between metal electrodes. This system exhibits enhanced heating at high biases and long phonon equilibration times. Despite the differences in their formulation, the static and dynamical methods capture local current-induced heating and inelastic corrections to the current with good agreement over a wide range of conditions, except in the limit of very high vibrational excitations, where differences begin to emerge.  
\end{abstract}
\maketitle

\section{Introduction}
The effects of inelastic interactions between current-carrying electrons and the vibrational motion of atomic nuclei is one of the principal phenomena of interest in the field of molecular electronics. These effects have been extensively studied experimentally in recent years \cite{agrait2003a,agrait2002a, smit, sakai}. The operation and properties of atomic-scale devices are strongly dependent on electron-ion interactions. The inelastic scattering of electrons by nuclei, and subsequent dynamical motion of the atoms, influences the transport properties of the device, while local Joule heating within the junction limits the stability of the device.

The simplest approach to such phenomena is lowest-order electron-phonon scattering theory, i.e. the Fermi Golden Rule (FGR). This includes the first order corrections to the electronic system from the electron-phonon interaction, which is treated as a perturbation. Phenomena such as the injection of power in the vibrational modes of atomic wires \cite{montgomery2003b} and corrections to the current-voltage spectrum which arise from the presence of inelastic electron-phonon scattering can be captured at a qualitative level within this framework. First-order perturbation theory however cannot be expected to handle the limit of strong electron-phonon coupling, or the effects of multiple scattering.

An established method of generalising the FGR to include higher-order processes is non-equilibrium Green's function theory (NEGF) \cite{haug1996, mahan}. One manner in which this is done is to consider only lowest-order Feynman diagrams in the expression for the self-energy and to expand the Dyson equation in a Born series in the free Green's functions. If the electronic Green's function used in the Dyson equation and in the calculation of the self-energy are the same, one obtains the Self-Consistent Born Approximation (SCBA). SCBA has been applied to inelastic transport both in model systems \cite{frederiksenMSc,galperin2004a,ryndyk2006a} and, together with first-principles electronic-structure calculations, in realistic atomic chains and molecular-wire systems \cite{frederiksen2004a,paulsson2005a,frederiksen2007a}. The SCBA technique is outlined further in the following section. The Green's function method can be applied also in the time domain, in order to take account of transient effects and the response of the system to dynamical driving fields \cite{jauho1994a,guo2005a}. 

Recently, an alternative method for inelastic transport has been proposed \cite{horsfield2004b, horsfield2005a, mceniry2007a} that differs from NEGF in philosophy and formulation. The key aim of this method is to  extend molecular dynamics by reinstating electron-nuclear correlations and the quantum nature of nuclei in order to produce a computationally tractable form of quantum correlated electron-ion dynamics (CEID) that retains inelastic electron-phonon interactions, energy transfer and dissipation between the two subsystems. Thus far, the method has been applied to inelastic $I-V$ spectroscopy in atomic wires \cite{horsfield2005a} and, when combined with electronic open boundaries, was used to calculate local heating in atomic wires, and its signature on the current, in real time \cite{mceniry2007a}. An outline of the method is given in Section \ref{section:ceid}.

In this paper, we report the first direct comparison of the dynamical CEID method with the SCBA. For this comparison, we have chosen a particular model system that exhibits interesting behaviour. We study a linear trimer weakly coupled to metal electrodes with the central atom in the trimer allowed to move. In the absence of electron-electron screening and coupling of vibrations to the surrounding lattice, this resonant system is found to undergo local Joule heating that is significantly larger than that obtained in a ballistic wire. The time taken for local phonons to equilibrate with the current-carrying electrons is also enhanced, and is strongly dependent on voltage.

The outline of the present work is thus as follows. In the next section, the SCBA formalism is outlined. Following that, the CEID methodology is outlined in Section \ref{section:ceid}, and it is shown, to lowest order in the electron-phonon coupling, that the steady-state solution to the one-particle electronic density matrix involves effective self-energies which are analogous to those in the Born Approximation. We also examine the infinite-mass limit of the CEID equations, and demonstrate that they reduce to the exact solution of a specific elastic scattering problem. The combination of CEID with electronic open boundaries is briefly summarized.

In Section \ref{section:applications}, the static and time-dependent methods are applied to our model system. The inelastic $I-V$ spectrum is analysed using both methods for two limiting regimes; one where (i) the moving ion is assumed to remain always in its ground state (the externally damped limit with perfect heat dissipation to the electrodes), and the other where (ii) no lattice heat conduction is allowed (the externally undamped limit with maximal heating). The inelastic current as a function of the thermal excitation of the quantum ion is studied for a variety of ionic masses, and the methods agree up to ionic vibrational energies $\sim$ 1 eV. Differences that emerge under more extreme conditions, and other directions for future work, are summarized at the end.

\section{The Self-Consistent Born Approximation (SCBA)} \label{section:scba}
In this section, the formalism of the SCBA is briefly outlined. The detailed description of the method is outlined elsewhere \cite{frederiksenMSc,galperin2004a,ryndyk2006a}. One assumes a coupled electron-phonon system within the harmonic approximation, whose Hamiltonian, in second quantization, is written as
\begin{eqnarray}
\hat{H} & = & \hat{H}_0 + \hat{H}_{\rm ph} + \hat{H}_{\rm e-ph}, \label{eq:scba_form} \\
\hat{H}_0 &= & \hat{H}_0 ( \{ c_k^{\dagger} \} ; \{ c_k \} ),\\
\hat{H}_{\rm ph} & = & \hbar \sum_{\lambda} \Omega_{\lambda} ( a_{\lambda}^{\dagger} a_{\lambda} + \frac{1}{2} ), \\
\hat{H}_{\rm e-ph} & = & \sum_{k,k'} \sum_{\lambda} M_{k, k'}^{\lambda} c_{k}^{\dagger} c_{k'} ( a_{\lambda}^{\dagger} + a_{\lambda} ) .
\end{eqnarray}
Here $\hat{H}_0$ is the electronic Hamiltonian described via the one-electron basis $\{ | k \rangle \}$, evaluated at the classical equilibrium nuclear positions $\{ R_0\}$, and $\{ c_{k}^{(\dagger)} \}$ is a corresponding set of one-electron annihilation (creation) operators. $\hat{H}_{\rm ph}$ is the phonon Hamiltonian for a set of uncoupled harmonic oscillators with $\{ a_{\lambda}^{(\dagger)} \}$ the set of annihilation (creation) operators within the occupation number representation, and $\Omega_{\lambda}$ is the vibrational frequency of mode $\lambda$. $\hat{H}_{\rm e-ph}$ describes the interaction between the electron and phonon subsystems, where the matrix $\hat{M}^{\lambda}$ is the electron-phonon coupling matrix for phonon mode $\lambda$. We also impose the non-crossing approximation, assuming that the interaction of the electron gas with the electron reservoirs is independent of its interaction with the vibrational modes of the system. 

We further assume that the electron Green's functions $\hat{G}_0^{+,\lessgtr}$ for the phonon-free electronic system can be evaluated. In the case of a nanoscale system coupled to external electronic reservoirs, these will explictly include the contribution due to the device-electrode coupling. The bare phonon Green's functions $D_0^{+,\lessgtr}$ in the frequency domain are those of a free harmonic oscillator of frequency $\Omega_0$,
\begin{eqnarray}
D_0^+ (\omega) & = &\frac{1}{\omega - \Omega_0  + \eye  \eta} - \frac{1}{\omega + \Omega_0 + \eye  \eta}, \\
D_0^\lessgtr (\omega) &=& - 2\pi \eye [ (N_{\rm ph} +1) \delta (\omega \pm \Omega_0) +N_{\rm ph} \delta ( \omega \mp \Omega_0)],
\end{eqnarray}
where $ \eta \to 0^+$ and $N_{\rm ph}$ is the phonon occupation number which in equilibrium is given by the Bose-Einstein distribution.

In the weak coupling limit, it is appropriate to consider only the lowest-order phonon contributions to the electron self-energy, i.e. to impose the Born Approximation (BA). Within the first Born Approximation, the self-energies are evaluated with the unperturbed Green's functions above and obtained by the Feynman rules as follows\footnote{Here we omit the so-called Hartree diagram since its contribution is frequency-independent, and it has no contribution to the inelastic signal in current-voltage spectra.}
\begin{eqnarray}
& & \hat{\Sigma}^{\lessgtr, {\rm BA}}_{\rm ph} (E)   = \frac{\eye}{2 \pi} \sum_{\lambda} \int \hat{M}^{\lambda} D_{0, \lambda}^{\lessgtr} (\omega) \hat{G}^{\lessgtr}_0 (E - \hbar \omega) \hat{M}^{\lambda} \diff \omega, \label{eq:ba_selfenergies1}\\
& &\hat{\Sigma}^{+, {\rm BA}}_{\rm ph} (E) = \frac{\eye}{2 \pi} \sum_{\lambda} \int \hat{M}^{\lambda} [ D_{0, \lambda}^< (\omega) \hat{G}^+_0 (E - \hbar \omega)  \nonumber \\
& & + D^+_{0,\lambda} (\omega) \hat{G}^<_0 (E - \hbar \omega) + D^+_{0, \lambda} (\omega) \hat{G}^+_0 (E - \hbar\omega)] \hat{M}^{\lambda} \diff \omega . \label{eq:ba_selfenergies2}
\end{eqnarray}
We neglect here the renormalization of the phonon modes due to the effect of the electrons which would appear via a self-energy analogous to those in Eqs.~(\ref{eq:ba_selfenergies1})-(\ref{eq:ba_selfenergies2}). This is appropriate when the mass of the ions is sufficiently large such that Migdal's theorem holds \cite{migdal}; however, the subsequent dispersion of the phonon Green's functions in energy space, which leads to a finite lifetime, cannot hence be taken into account. To go beyond the BA one can perform a self-consistent procedure for the electronic Green's functions such that the Green's function which satisfies the Dyson and Keldysh equations and that used to evaluate Eqs.~(\ref{eq:ba_selfenergies1})-(\ref{eq:ba_selfenergies2}) are equivalent. This procedure is known as the self-consistent Born Approximation (SCBA).

The self-consistent Green's functions thus obtained may be used to calculate properties of interest, such as the steady-state inelastic current and the power injected from the electrons into the vibrational modes of the system\cite{frederiksenMSc,galperin2004a,ryndyk2006a,meir1992a}.

\section{Correlated Electron-Ion Dynamics (CEID)} \label{section:ceid}
\subsection{Formulation}\label{section:ceid:formulation}
One of the advantages of using dynamical methods as a basis for electronic transport calculations is that the interplay between  electrical properties and atomic motion can be addressed. Conventional Born-Oppenheimer molecular dynamics simulations enable the calculation of current-induced corrections to atomic forces. However, in such simulations the scattering of electrons from ions is purely elastic and the electronic structure for a given ionic geometry remains in a steady state.

A principal goal, therefore, of extending molecular dynamics beyond the adiabatic approximation is to understand phenomena in which both the electrons and ions depart from equilibrium, the subsequent interactions between them, and the exchange of energy between the two subsystems.  Correlated electron-ion dynamics constitutes an attempt to introduce correlated electron-ion fluctuations as low-order corrections to Ehrenfest dynamics, by expanding the electron-ion quantum Liouville equation in powers of such fluctuations \cite{horsfield2004b, horsfield2005a}. The method enables the description of the energy exchange between electrons and ions in a non-equilibrium environment and the dynamical response of the electron gas to the variations in the ionic distribution.

The idea of the method is best illustrated by applying it to a Hamiltonian of the form (\ref{eq:scba_form}) (now expressed in first quantization), in which we consider electrons linearly coupled to a single harmonic oscillator,
\begin{eqnarray}
\hat{H}_{\rm eI}& =& \hat{H}_0^{(N_{\rm e})} - \hat{F}^{(N_{\rm e})} \cdot (\hat{R} - R_0) + \frac{\hat{P}^2}{2M} \\ \nonumber
& & + \frac{1}{2} K^{\rm BO} (\hat{R} - R_0)^2  \nonumber \\ 
& =& \hat{H}_{\rm e}^{(N_{\rm e})} (\hat{R}) + \frac{\hat{P}^2}{2M} + \frac{1}{2} K^{\rm BO} (\hat{R}- R_0)^2 \label{eq:eIHam}
\end{eqnarray}
where $\hat{H}_0^{(N_{\rm e})} = \hat{H}_{\rm e}^{(N_{\rm e})} (R_0)$ is the $N_{\rm e}$-electron Hamiltonian in the presence of a classical oscillator centered at the equilibrium position $R_0$ and $\hat{F}^{(N_{\rm e})}$ denotes the electron-ion coupling operator, $\hat{P}, \hat{R}$ are respectively the ionic momentum and position operators, and $K^{\rm BO}$ is the Born-Oppenheimer spring constant of the harmonic oscillator. The combined electron-ion density matrix $\hat{\rho}_{\rm eI}$ satisfies the quantum Liouville equation:
\begin{displaymath}
\dot{\hat{\rho}}_{\rm eI} = \frac{1}{\eye \hbar} [ \hat{H}_{\rm eI} , \hat{\rho}_{\rm eI} ].
\end{displaymath}

Tracing over ionic degrees of freedom leads to the following set of coupled equations of motion:
\begin{eqnarray}
\dot{\hat{\rho}}_{\rm e}^{(N_{\rm e})} & =& \frac{1}{\eye \hbar} [ \hat{H}_{\rm e}^{(N_{\rm e})} (\bar{R}) , \hat{\rho}_{\rm e}^{(N_{\rm e})} ] - \frac{1}{\eye \hbar} [ \hat{F}^{(N_{\rm e})}, \hat{\mu}^{(N_{\rm e})}], \label{eq:rho}\\
\dot{\hat{\mu}}^{(N_{\rm e})} &=& \frac{1}{\eye \hbar} [ \hat{H}_{\rm e}^{(N_{\rm e})} (\bar{R}), \hat{\mu}^{(N_{\rm e})}] - \frac{1}{\eye \hbar} [ \hat{F}^{(N_{\rm e})} , \hat{\mu}_2^{(N_{\rm e})} ] + \frac{\hat{\lambda}^{(N_{\rm e})}}{M}, \label{eq:mu} \\
\dot{\hat{\lambda}}^{(N_{\rm e})} &=& \frac{1}{\eye \hbar} [ \hat{H}_{\rm e}^{(N_{\rm e})} (\bar{R}), \hat{\lambda}^{(N_{\rm e})}] - \frac{1}{\eye \hbar} [ \hat{F}^{(N_{\rm e})}, \hat{\chi}_2^{(N_{\rm e})} ] \nonumber \\
& & + \frac{1}{2} \{ \Delta \hat{F}^{(N_{\rm e})}, \hat{\rho}_{\rm e}^{(N_{\rm e})} \} - K^{\rm BO} \hat{\mu}^{(N_{\rm e})}, \label{eq:lambda} \\
\dot{\mu}_2^{(N_{\rm e})}& =& \cdots ,\label{eq:mu2}
\end{eqnarray}
where $\bar{R} = \bar{R} (t)$ is the classical mean trajectory of the oscillator. $\hat{\rho}_{\rm e}^{(N_{\rm e})} = \Tr_{\rm I} \{ \hat{\rho}_{\rm eI}\}$ is the $N_{\rm e}$-particle electronic density matrix,  $\hat{\mu}^{(N_{\rm e})} = \Tr_{\rm I} \{ \Delta \hat{R} \hat{\rho}_{\rm eI}\}$, $\hat{\lambda}^{(N_{\rm e})} = \Tr_{\rm I} \{ \Delta \hat{P} \hat{\rho}_{\rm eI}\}$, $\Delta \hat{R} = \hat{R} - \bar{R}$, $\Delta {\hat{P}} = \hat{P} - \bar{P}$, $\Delta{\hat{F}}^{(N_{\rm e})} = \hat{F}^{(N_{\rm e})} - \Tr \{ \hat{\rho}_{\rm e}^{(N_{\rm e})} \hat{F}^{(N_{\rm e})} \}$, $\bar{F} = \dot{\bar{P}} = \Tr \{ \hat{\rho}_{\rm e}^{(N_{\rm e})} \hat{F}^{(N_{\rm e})} \} - K^{\rm BO} ( \bar{R} - R_0)$, $ \bar{P} = M \dot{\bar{R}}$.
 
In order to obtain a closed set of equations, the right-hand sides above are truncated to lowest non-trivial order in the electron-ion coupling. Thus, in the second term in (\ref{eq:mu}) and in the second term in (\ref{eq:lambda}), we make the mean-field approximations
\begin{eqnarray}
\hat{\mu}_2^{(N_{\rm e})} &=& \Tr_{\rm I} \{ ( \Delta \hat{R})^2 \hat{\rho}_{\rm eI} \} \approx \langle ( \Delta \hat{R} )^2 \rangle \hat{\rho}_{\rm e}^{(N_{\rm e})} = C_{RR} \hat{\rho}_{\rm e}^{(N_{\rm e})}, \label{eq:crr}\\
\hat{\chi}_2^{(N_{\rm e})} &=& \frac{1}{2} \Tr_{\rm I} \{ \{ \Delta \hat{R} , \Delta \hat{P} \} \hat{\rho}_{\rm eI} \} \approx \frac{1}{2} \langle \{ \Delta \hat{R}, \Delta \hat{P} \} \rangle \hat{\rho}_{\rm e}^{(N_{\rm e})} \nonumber \\
& &  = C_{RP} \hat{\rho}_{\rm e}^{(N_{\rm e})}.\label{eq:cpr}
\end{eqnarray}
The above equations are many-electron equations of motion; these are reduced to one-electron form by tracing out all other electrons with the help of a Hartree-Fock approximation to the two-electron density matrix. This procedure is described in detail elsewhere \cite{horsfield2005a} and leads to the one-electron equations of motion
\begin{eqnarray}
\dot{\hat{\rho}}_{\rm e} & =& \frac{1}{\eye \hbar} [ \hat{H}_{\rm e} (\bar{R}), \hat{\rho}_{\rm e}] - \frac{1}{\eye \hbar} [ \hat{F} , \hat{\mu}], \label{linearised_equations1}\\
\dot{\hat{\mu}} & =& \frac{1}{\eye \hbar} [ \hat{H}_{\rm e} (\bar{R}), \hat{\mu}] - \frac{1}{\eye \hbar} C_{RR} [ \hat{F} , \hat{\rho}_{\rm e}] + \frac{\hat{\lambda}}{M}, \label{eq:linear_mu}\\
\dot{\hat{\lambda}} & =& \frac{1}{\eye \hbar} [ \hat{H}_{\rm e} (\bar{R}), \hat{\lambda}] + \frac{1}{2} ( \hat{F} \hat{\rho}_{\rm e} + \hat{\rho}_{\rm e} \hat{F} ) - \hat{\rho}_{\rm e} \hat{F} \hat{\rho}_{\rm e} \nonumber \\
& & - \frac{1}{\eye \hbar} C_{RP} [ \hat{F}, \hat{\rho}_{\rm e}] - K^{\rm BO} \hat{\mu}, \label{linearised_equations2}
\end{eqnarray}
where all operators are now one-electron operators, and where we define
\begin{eqnarray*}
\hat{\rho}_{\rm e} = N_{\rm e} \Tr_{ {\rm e}, 2\ldots N_{\rm e}} \hat{\rho}_{\rm e}^{(N_{\rm e})}, \\ 
\hat{\mu} = N_{\rm e} \Tr_{{\rm e}, 2\ldots N_{\rm e}} \hat{\mu}^{(N_{\rm e})}, \\
\hat{\lambda} = N_{\rm e} \Tr_{{\rm e}, 2\ldots N_{\rm e}} \hat{\lambda}^{(N_{\rm e})} .
\end{eqnarray*}

\subsection{Weak scattering limit}
In this section, we make an approximate, although revealing, connection between the steady-state limit of the CEID equations above and the SCBA. We assume that the vibration is in an oscillator eigenstate with $N_{\rm ph}$ phonons, where $K^{\rm BO} C_{RR} = (N_{\rm ph} + \frac{1}{2}) \hbar \Omega_0$ and $C_{RP} = 0$. We imagine that the phonon-free electron system has settled in a steady-state with a one-electron density matrix $\hat{\rho}_{\rm e} = \int \hat{\rho}_{\rm e} (E) \diff E$, where $\hat{\rho}_{\rm e} (E) = \sum_{\alpha} |\alpha \rangle f_{\alpha} \delta ( E - E_{\alpha}) \langle \alpha |$ is the energy-resolved density matrix. For an infinite open current-carrying system, the one-electron states $|\alpha \rangle$ with energies $E_{\alpha}$ would be Lippmann-Schwinger scattering wavefunctions with occupancies $f_{\alpha}$ set by the battery terminals. We ignore variations in $\bar{R}$ relative to the equilibrium position $R_0$ (hence $\hat{H}_{\rm e} (\bar{R}) = \hat{H}_0$ above), and we can then solve (\ref{eq:linear_mu})-(\ref{linearised_equations2}) to lowest order in $\hat{F}$. This is done in \cite{horsfield2005a}. Taking the long-time limit of the result for $\hat{\mu}$ gives 
\begin{widetext} 
\begin{eqnarray}
\mu_{\alpha \beta}& =& - F_{\alpha \beta} \frac{\hbar}{2 M \Omega_0} \left [ f_{\alpha} ( 1 - f_{\beta}) \left ( \frac{N_{\rm ph}}{E_{\alpha} - E_{\beta} + \hbar \Omega_0 - \eye \epsilon} +  \frac{N_{\rm ph} + 1}{E_{\alpha} - E_{\beta} - \hbar \Omega_0 - \eye \epsilon} \right ) \right. \nonumber \\ 
&&- \left.  f_{\beta} ( 1- f_{\alpha}) \left ( \frac{N_{\rm ph} + 1}{E_{\alpha} - E_{\beta} + \hbar \Omega_0 - \eye \epsilon} + \frac{N_{\rm ph}}{E_{\alpha} - E_{\beta} - \hbar \Omega_0 - \eye \epsilon} \right ) \right ].\nonumber\\
\end{eqnarray}
\end{widetext}
where $\epsilon \to 0^+$ and $\Omega_0^2 = K^{\rm BO}/M$.  This expression may be substituted into Eq (\ref{linearised_equations1}). The commutator $[ \hat{F}, \hat{\mu}]$, which describes the electron-phonon coupling, then becomes: 
\begin{eqnarray}
[ \hat{F}, \hat{\mu}] &=& \int [ \hat{\rho}_{\rm e} (E) \hat{\Sigma}_{\rm ph}^- (E) - {\rm h.c.} ] \diff E \nonumber \\
& & - \frac{1}{2 \pi \eye} \int [ \hat{\Sigma}_{\rm ph}^< (E) \hat{G}_0^- (E) - {\rm h.c.} ] \diff E,
\end{eqnarray}
where $\hat{G}_0^-(E)$ is the advanced phonon-free electronic Green's function and the self-energies $\hat{\Sigma}_{\rm ph}^{\pm, <} (E)$ are given by
\begin{widetext}
\begin{eqnarray}
\hat{\Sigma}^{\pm}_{\rm ph} (E) & =& \frac{\hbar}{2 M \Omega_0} \sum_{\alpha} \hat{F} | \alpha \rangle \left ( \frac{(N_{\rm ph} + 1)(1 - f_{\alpha})}{E - E_{\alpha} - \hbar \Omega_0 \pm \eye \epsilon} + \frac{(N_{\rm ph} + 1) f_{\alpha}}{E - E_{\alpha} + \hbar \Omega_0 \pm \eye \epsilon} + \frac{N_{\rm ph} (1 - f_{\alpha} )}{E - E_{\alpha} + \hbar \Omega_0 \pm \eye \epsilon} + \frac{N_{\rm ph} f_{\alpha} }{ E - E_{\alpha} - \hbar \Omega_0 \pm \eye \epsilon} \right ) \langle \alpha | \hat{F}, \nonumber \\ 
\hat{\Sigma}^<_{\rm ph} (E) &=& 2 \pi \eye \frac{\hbar}{2 M \Omega_0} \sum_{\alpha} \hat{F} | \alpha \rangle  \left [ (N_{\rm ph} + 1) \delta ( E+ \hbar \Omega_0 - E_{\alpha} ) + N_{\rm ph} \delta ( E- \hbar \Omega_0 - E_{\alpha}) \right ] f_{\alpha}\langle \alpha | \hat{F}. \nonumber\\ \label{eq:phonon_self_energies}
\end{eqnarray}
\end{widetext}
It is shown in Appendix \ref{app:baself} that these expressions for the self-energies are the same as those in the first Born approximation. We have thus established that in the limit of weak electron-ion coupling, the CEID and SCBA steady states agree.

\subsection{Large Mass Limit in CEID} \label{section:large_mass}
In this section, we examine the limit of infinite mass, in which Eqs. (\ref{linearised_equations1}-\ref{linearised_equations2}) can again be solved analytically. In that case $\bar{R} = \bar{R} (0)= R_0$ is a constant, the equations of motion for $\hat{\mu}$ and $\hat{\lambda}$ decouple, and (\ref{linearised_equations1}-\ref{eq:linear_mu}) reduce to
\begin{eqnarray}
\eye \hbar \dot{\hat{\rho}}_{\rm e} & =& [ \hat{H}_0, \hat{\rho}_{\rm e}] - [ \hat{F}, \hat{\mu} ] \nonumber \\
\eye \hbar \dot{\hat{\mu}} & = & [ \hat{H}_0, \hat{\mu} ] - C_{RR} [\hat{F}, \hat{\rho}_{\rm e}]. \label{eq:ceid_big_mass}
\end{eqnarray}
To derive this we now examine the following {\it elastic} scattering problem. We consider non-interacting electrons coupled linearly to an infinitely heavy classical degree of freedom $X$, with some time-independent statistical distribution  $\chi (X)$. We have the one-electron Hamiltonian
\begin{equation}
\hat{H} (X)=  \hat{H}_0 - \hat{F} X .\label{eq:elastic_HX}
\end{equation}
Imagine solving the Liouville equation $\eye \hbar \dot{\hat{\rho}} (X,t) = [ \hat{H}(X), \hat{\rho} (X,t)]$ for the one-electron density matrix $\hat{\rho} (X,t)$. Define
\begin{eqnarray}
\hat{\rho}_{\rm e} (t)& =& \int \hat{\rho}(X, t) \chi (X) \diff X \\
\hat{\mu} (t) & = & \int X \hat{\rho} (X, t) \chi (X) \diff X \\
\hat{\mu}_2 (t) & = & \int X^2 \hat{\rho} (X, t) \chi(X) \diff X .
\end{eqnarray}
Then, 
\begin{eqnarray}
\eye \hbar \dot{\hat{\rho}}_{\rm e} & = & [ \hat{H}_0 , \hat{\rho}_{\rm e}] - [ \hat{F}, \hat{\mu}] \nonumber \\
\eye \hbar \dot{\hat{\mu}} & = & [ \hat{H}_0, \hat{\mu}] - [\hat{F}, \hat{\mu}_2]. \label{eq:big_mass}
\end{eqnarray}
Now consider the distribution
\begin{equation}
\chi (X) = \frac{1}{2} [ \delta ( X - a) + \delta (X + a)]. \label{eq:chi_x}
\end{equation}
We then have \underline{exactly}
\begin{displaymath}
\hat{\mu}_2 = C_{RR} \hat{\rho}_{\rm e}; \ C_{RR} = \int X^2 \chi (X) \diff X = a^2,
\end{displaymath}
and equations (\ref{eq:big_mass}) reduce identically to (\ref{eq:ceid_big_mass}). Therefore, in the large-mass limit, CEID is algebraically equivalent to the elastic scattering problem defined by Eqs. (\ref{eq:elastic_HX}) and (\ref{eq:chi_x}). This equivalence will be used later to benchmark the approximate OB method used in CEID. 

\subsection{General CEID Equations} \label{section:general_ceid}
The original formulation of CEID, which will be used for the calculations in Section \ref{section:applications}, starts from a more general Hamiltonian than that in equation (\ref{eq:eIHam}). We start formally from the full electron-nuclear Hamiltonian $\hat{H}_{\rm eI}$, which we partition as in Section \ref{section:ceid:formulation}, $\hat{H}_{\rm eI} = \hat{H}_{\rm e}^{(N_e)} (\hat{R}) + \hat{T}_I + \hat{H}_I (\hat{R})$, where $\hat{H}_{\rm e}^{(N_e)} (\hat{R})$ includes the bare electron-ion interaction, $\hat{T}_{\rm I}$ is the nuclear kinetic energy operator, and $\hat{H}_I$ is the bare ion-ion interaction potential. Within the weak-coupling approximation considered here, this Hamiltonian is expanded about the mean ionic trajectory $\bar{R}$ to second order in $\Delta \hat{R}$.
\begin{eqnarray}
\hat{H}_{\rm eI} &\approx &\hat{H}_{\rm e}^{(N_{\rm e})} (\bar{R}) + \hat{H}_I (\bar{R}) - ( \hat{F}^{(N_{\rm e})} ( \bar{R}) + F_{\rm I} (\bar{R})) \cdot \Delta \hat{R} \nonumber \\
& & + \frac{1}{2} ( \hat{K}^{(N_{\rm e})} (\bar{R}) + K_{\rm I} (\bar{R})) (\Delta \hat{R})^2 + \hat{T}_{\rm I}, 
\end{eqnarray}
where $\hat{K}^{(N_{\rm e})} (\bar{R}) = \partial^2 \hat{H}_{\rm e}^{(N_{\rm e})} (\bar{R})/\partial \bar{R}^2$, $F_{\rm I} (\bar{R}) = -\partial \hat{H}_{\rm I} (\bar{R})/\partial \bar{R}$, and $K_{\rm I} (\bar{R}) = \partial^2 \hat{H}_{\rm I} (\bar{R})/\partial \bar{R}^2$. This Hamiltonian is inserted into the full quantum Liouville equation, and an analogous procedure to that which led to Eqs. (\ref{linearised_equations1}-\ref{linearised_equations2}) is undertaken. The reduction of the equations of motion to one-electron form requires an extension to the Hartree-Fock approximation to the two-electron density matrix \cite{horsfield2005a} to allow for the essential non-idempotency introduced by electron-ion correlations. This is necessary in order to take account of the screening of the bare ion-ion interaction by the electron-ion interaction and the corresponding contributions to the effective stiffness. Since the CEID equations are derived from the bare interaction potentials in the system, the effective stiffnesses, phonon modes and frequencies are no longer an input, but are generated as part of the simulation. Furthermore, the inclusion of second-order electron-ion coupling (via $\hat{K}$) arises naturally from the second-order expansion. A reformulation of the CEID expansion for systems with strong electron-nuclear correlations is developed in \cite{stella2007a}.  The full set of one-electron equations of motion, including equations of motion for  the ionic variables $\bar{R}, \bar{P}, C_{RR}, C_{RP}$, and $C_{PP} = \langle ( \Delta \hat{P} )^2 \rangle$, are reproduced in Appendix \ref{app:ceidequations}.

\subsection{CEID with Open Boundaries (OB)} \label{section:ceid_plus_ob}
The CEID calculations below use the open-boundary method described in \cite{mceniry2007a}. We consider a finite, though possibly large, system $S=LCR$ consisting of electrodes $L$ and $R$ with a region $C$ between them. All dynamical scattering is assumed to be confined to $C$. Each finite electrode is embedded in, and weakly coupled to, a sea of external probes $P$. Probes coupled to $L(R)$ are maintained at electrochemical potential $\mu_{L(R)}$, with corresponding Fermi-Dirac distributions $f_{L(R)} (E)$. 
The open-boundary equations of motion for the one-electron operators $\hat{\rho}_{\rm e} , \hat{\mu}, \hat{\lambda}$ in $S$ are 
\begin{equation}
\eye \hbar \dot{\hat{q}} = [ \hat{H}_{\rm e}, \hat{q} ] + \hat{\Lambda}^{(q)} + \hat{D}^{(q)} \quad \hat{q} = \hat{\rho}_{\rm e}, \hat{\mu}, \hat{\lambda}, \label{eq:summary}
\end{equation}
where $\hat{\Lambda}^{(q)}$ denotes the electron-ion dynamical scattering terms, and $\hat{D}^{(q)}$ denotes the open-boundary driving terms. These driving terms are
\begin{eqnarray}
\hat{D}^{(\rho_{\rm e})} & =& \hat{\Sigma}^+ \hat{\rho}_{\rm e} - \hat{\rho}_{\rm e} \hat{\Sigma}^- \nonumber \\
& &  + \int [ \hat{\Sigma}^< (E) \bar{G}^- (E) - \bar{G}^+ (E) \hat{\Sigma}^< (E) ] \diff E, \label{eq:drho} \\
\hat{D}^{(\mu)} & = & \hat{\Sigma}^+ \hat{\mu} - \hat{\mu} \hat{\Sigma}^-, \label{eq:dmu} \\
\hat{D}^{(\lambda)} & = & \hat{\Sigma}^+ \hat{\lambda} - \hat{\lambda} \hat{\Sigma}^-, \label{eq:dlambda}
\end{eqnarray}
where
\begin{eqnarray}
\hat{\Sigma}^{\pm} & = &\mp \eye \frac{\Gamma}{2} \hat{1}_L \mp \eye \frac{\Gamma}{2} \hat{1}_R, \label{eq:sigmapm} \\
\hat{\Sigma}^< (E) & = &\frac{\Gamma}{2 \pi} f_L (E) \hat{1}_L + \frac{\Gamma}{2 \pi} f_R (E) \hat{1}_R, \label{eq:sigmaless} \\
\bar{G}^{\pm} (E) & = & ( E  - \hat{H}_0 - \hat{\Sigma}^{\pm} \pm \eye \Delta )^{-1}, \label{eq:gbar}
\end{eqnarray}
where $\hat{H}_0={\hat H}_0 (R_0)$ is the phonon-free one-electron Hamiltonian and $\hat{1}_{M}$ denotes the identity operator in region $M$.

These equations are obtained by making two approximations. The first is to take the wide-band limit in the external probes $P$. This makes the $SP$ coupling strength, $\Gamma$, an energy-independent parameter and the extraction terms (the first two terms in Eq. (\ref{eq:drho})) temporally local. The second approximation is the introduction of a dephasing mechanism in the $SP$ coupling, characterized by an energy scale, $\Delta$, and a dephasing time $\tau_{\Delta} = \hbar / \Delta$. Provided $C$ is long enough, so that $\tau_{\Delta}$ is less than the time for signals to travel between $L(R)$ and $C$, the dephasing mechanism breaks the coherence between injection into $L(R)$ and subsequent scattering in $C$. This in turn has the effect of making the injection terms (the second two terms in Eq. (\ref{eq:drho})) independent of the dynamical scattering in $C$. Otherwise, the Green's function in the injection terms would contain a self-energy describing the scattering in $C$.

The resultant open-boundary scheme has the benefit of being temporally local. However, the cost is that the dephasing mechanism above is in turn equivalent to replacing the true Fermi-Dirac distributions in the probes $P$ by effective distributions with an energy broadening $\sim 2\Delta$, resulting in a corresponding loss of energy resolution. The longer the device $C$, the smaller the broadening, required to mask the dynamical scattering in $C$. In the calculations presented here, we have set $\Delta = 0$. The resulting injection terms differ from those generated by the value of $\Delta$, appropriate for a given device length, by an energy uncertainty that itself disappears with $\Delta$. In the absence of phonons, $\Delta = 0$ generates the exact unbroadened elastic steady-state solution for the multiple probe battery, which in turn gives arbitrarily close approximations to the conventional two-terminal Landauer steady state \cite{mceniry2007a}.

\begin{figure}
\includegraphics[totalheight = 0.25\textheight,clip=]{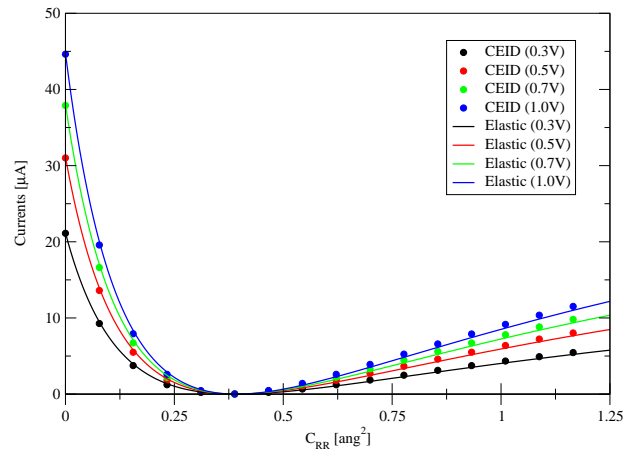}
\caption{Steady-state currents as a function of effective cross section for a single degree of freedom of infinite mass, for a variety of biases, in the CEID approach. The device length used here was 401 atoms, with 100 atoms in each electrode, with the parameters $\Gamma$ and $\Delta$ set to $0.4$ eV and $0.0$ eV respectively. The point at which the current drops to zero corresponds to one of the hopping integrals in the Hamiltonian (\ref{eq:elastic_HX}) going to zero. (Color online)} \label{fig:fake_ceid}
\end{figure}

The OB method is tested by applying it to CEID in the large-mass limit considered in Section \ref{section:large_mass}. We take the electronic system to be a resonant trimer, described in more detail in the following section, within a $1s$ tight-binding model with non-interacting electrons. The results obtained from the CEID calculations are compared to the exact {\it static} elastic steady-state, which can be calculated separately,  within numerical precision, from the Landauer formalism, and which, in the absence of any approximation in the OBs, must agree identically with the large-mass CEID steady-state. The two steady-state currents as a function of effective cross-section $C_{RR}$ are presented in Fig.~\ref{fig:fake_ceid} for a variety of biases, with excellent agreement. 

\section{Results} \label{section:applications}
An electron within a resonant molecule characterised by an energy width $\delta E$ will have a lifetime $t \sim \hbar/\delta E$. If this lifetime is sufficiently large, the electron may be expected to undergo several electron-phonon interactions, which may lead to high excitation of the vibrational modes of the molecule. Such multiple electron-phonon scattering events lie beyond lowest-order perturbation theory. However, since the SCBA effectively sums the low-order scattering events to infinite order, it will capture at least some of the pertinent physics. In view of the equivalence of CEID and the SCBA in the weak-coupling limit established  in the previous section, we  conjecture that the CEID equations will be able to capture the phenomena of interest. Neither calculation can be expected to be correct in the limit of strong electron-phonon coupling, or in the limit of high phonon excitation (in which case, effects of anharmonicity would also be significant).

\begin{figure}
\includegraphics[totalheight = 0.1\textheight,clip=]{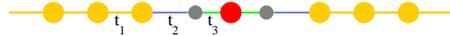}
\caption{Linear trimer weakly coupled to two one-dimensional electrodes. In the simulations considered here, only the central atom of the trimer is allowed to undergo vibrational motion. $t_1$ is the nearest-neighbour hopping matrix element in the metal electrode, $t_2$ is the electrode-trimer hopping integral, and $t_3$ is the intra-trimer hopping integral. (Color online)} \label{fig:application}
\end{figure}

In this section, we compare SCBA and CEID for the following model resonant system. We consider a linear trimer, illustrated in Fig.~\ref{fig:application}, which is weakly coupled to two one-dimensional perfect metal electrodes. The central atom of the trimer is treated as a dynamical quantum ion, allowed to move longitudinally. We assume non-interacting electrons throughout.

Due to the absence of electron-electron screening, only electron-ion interactions are present in the Hamiltonian; these are described via a single-orbital tight-binding model. In the CEID simulations, the ion-ion interactions (which are required to calculate the dynamical matrix) are described through a repulsive pair potential, with both sets of parameters fitted to bulk gold \cite{stch}. The bond length in the electrode is 2.5 {\AA} which corresponds to a hopping matrix elements of $t_1 \sim -3.88$ eV. The electrode-trimer distance is 3.509 {\AA}, corresponding to a hopping integral of $t_2\sim -1.00$ eV, while the intra-trimer bond length is also 2.5 {\AA} (i.e., $t_3 \sim -3.88$ eV). All onsite energies are set to zero. The electron-phonon coupling matrix $\hat{M}$ used in the SCBA calculations is chosen to be $\hat{M} =  -\sqrt{ \hbar/ 2 M \Omega_0} \hat{F} (R_0)$, derived from the same TB model as that used in CEID. The CEID calculations have been carried out with our parallel computer code pDINAMO \cite{capability}, an implementation of the CEID formalism developed to run on massively parallel computers.

With the present parameters and a band-filling of $0.5$, a resonance of width $\sim 0.54$ eV centred at the Fermi energy appears in the elastic transmission function. Based on considerations of the electron Fermi velocity and the geometry of the resonance, for the ionic mass considered below, we expect multiple electron-phonon interactions in the time interval corresponding to this width.

\subsection{Inelastic $I-V$ characteristics in the externally damped limit}
The first comparison between the two methods is to calculate the low-temperature inelastic correction to the current-voltage spectrum for the trimer in the externally damped limit. We assign a mass of 1 atomic mass unit (amu) to the moving atom, such that its Born-Oppenheimer vibrational frequency $\Omega_0$ is $\hbar \Omega_0 \sim 0.20$ eV. In the OB CEID calculations, the total number of atoms in the chain is 601, with 100 assigned to each electrode, with the probe-electrode coupling $\Gamma = 0.4$ eV. The second-order variables $C_{RR}, C_{PP}$ are set to those of the vibrational ground state, and kept ``frozen'' throughout the simulation. This, therefore, corresponds to the limit of perfect dissipation of energy away from the phonon modes. In the SCBA calculations, the occupation number of the phonon mode was effectively kept at $N_{\rm ph}\sim 0$.

\begin{figure}
\centering
\includegraphics[totalheight = 0.25\textheight,clip=]{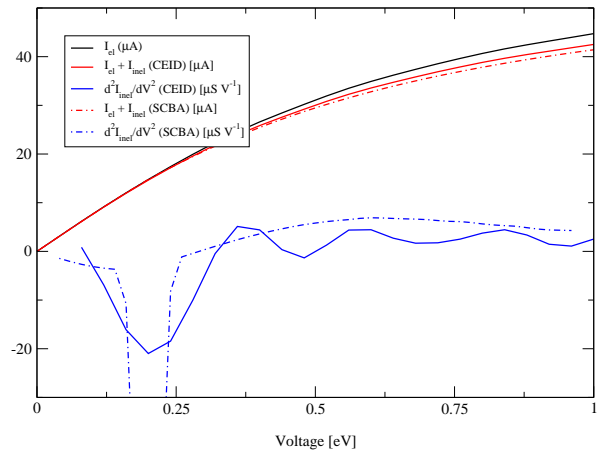}
\caption{Current-voltage spectrum and second derivative of inelastic current for the trimer system in the externally damped limit. (Color online) } \label{fig:ivspectrum}
\end{figure}

The current-voltage spectra obtained for the two methods are shown in Fig.~\ref{fig:ivspectrum}, together with the second derivative of the inelastic contribution to the current. Both methods capture the inelastic feature at the correct frequency and the overall drop in the conductance is similar. The feature obtained using the CEID calculations is rather broad; this results from the absence of an effective phonon contribution to the electron self-energy in the OB formalism, and from finite size effects (since the energy levels of the system are discrete). The width of the SCBA feature is a result of finite electron temperature as well as the numerical procedure for obtaining the second derivative.

\subsection{Heating and Equilibration}
The trimer system explored here exhibits two notable characteristics with regard to the heating of its vibrational mode. The effective phonon occupancy obtained at high voltages ($V \gg \hbar \Omega_0$) is significantly higher than that obtained in a ballistic atomic chain. In addition, the time taken for the vibrations to equilibrate with the electron gas is long\footnote{At $V = 1$V, the equilibration time is approximately $200$ fs -- about 4 times that of a ion of the same mass in a ballistic chain. Furthermore, within the FGR, the scaling of this time is linear with mass.}. The origin of these properties lies in the resonant character of the system, and can be understood,  at least qualitatively, within the FGR (see Appendix \ref{appendix:fgr}).

\begin{figure}
\centering
\includegraphics[width = \columnwidth,clip=]{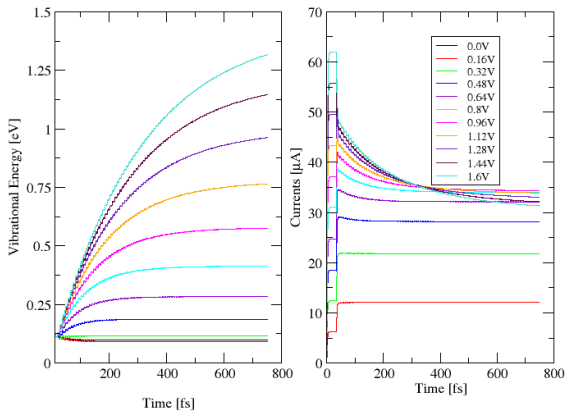}
\caption{Total vibrational energy of a single dynamical ion as a function of time and corresponding electronic current, for various electrochemical potential differences in the CEID approach. (Color online)} \label{fig:current_heating}
\end{figure}

To simulate heating in the CEID calculations, we ``unfreeze'' the phonon modes and allow the vibrational degrees of freedom to respond to the current-carrying electronic structure. We take the total vibrational energy to be $C_{PP}/M$, which assumes equipartitioning of the vibrational energy between the kinetic and potential energies and includes the zero-point energy\footnote{Strictly speaking, the ionic vibrational energy should also include a contribution from the classical kinetic energy, $\bar{P}^2/2M$, of the centroid, but in the present simulations, that contribution is negligible.}. Fig.~\ref{fig:current_heating} shows the  current and total vibrational energy as a function of time for various applied voltages. At voltages below the inelastic threshold the total vibrational energy remains flat. Above the inelastic threshold, as the voltage increases, we see the two features of the heating mentioned above; the total energy increases greatly and the time taken for equilibration also increases. Furthermore, at high voltages, the current traces ``cross over'', an indication of the onset of negative differential resistance.

\begin{figure}
\centering
\includegraphics[totalheight = 0.25\textheight,clip=]{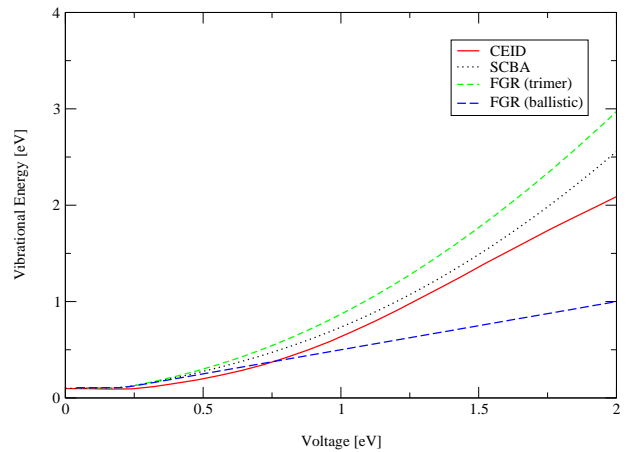}
\caption{Maximal vibrational energy as a function of voltage for the resonant trimer system, in the dynamical CEID calculation, the SCBA and lowest-order perturbation theory. Also shown is the vibrational energy obtained within the FGR for a single ion (with the same angular frequency) in a ballistic chain. (Color online)} \label{fig:heating}
\end{figure}

In order to compare the results here with the SCBA, we extrapolate the vibrational energy as a function of voltage to infinite times, and compare those with the maximal vibrational energy obtained in the SCBA, in the undamped limit. These results are illustrated in Fig.~\ref{fig:heating}, together with the maximal heating according to the FGR for the present system, and for a quantum ion of the same frequency in a ballistic chain. It is seen that the SCBA and CEID are in good agreement up to very high voltages; for the highest voltage on the plot, the effective phonon occupancy is $N_{\rm ph} \sim 10$. In both cases, the maximal vibrational energy significantly exceeds that for the ballistic chain.  The FGR calculation on the present system deviates from the other methods at high bias, indicating that CEID and the SCBA are capturing higher-order processes that are absent from the lowest-order perturbative treatment.  There are two regions of disagreement between CEID and the SCBA; at voltages just above the inelastic threshold the heating obtained from the CEID calculation is lower than that in the SCBA or FGR calculations. We speculate that this is due at least in part to the width of the inelastic spectral feature introduced by the inexact OB method used in the CEID calculation; the full effect of the oscillator is gradually seen with increasing bias. Secondly, at very high voltages, the increase in vibrational energy in the CEID calculations tapers off, which may be due to the explicit inclusion of second-order electron-ion coupling. 

\begin{figure}
\centering
\includegraphics[totalheight = 0.25\textheight,clip=]{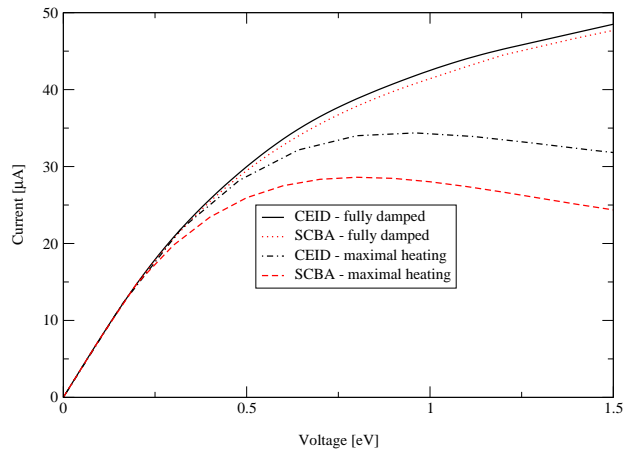}
\caption{Asymptotic values of the current as a function of voltage for the resonant trimer system, in the maximally-heated limit, for the CEID and SCBA methods. Also illustrated are the steady-state currents in the fully damped limit, illustrating the effect of the heating. (Color online)} \label{fig:fully_heated_iv}
\end{figure}

As a further comparison of CEID and the SCBA, we can consider the asymptotic values of the inelastic currents in the maximally-heated limit. These are presented in Fig.~\ref{fig:fully_heated_iv} and again there is good qualitative agreement between the two methods. In particular, both methods demonstrate that negative differential resistance will occur in this system in the undamped limit, although the methods predict a slightly diffferent value for the voltage at which the maximum current is achieved.

\subsection{Inelastic current as a function of cross-section and ionic mass}
\begin{figure}
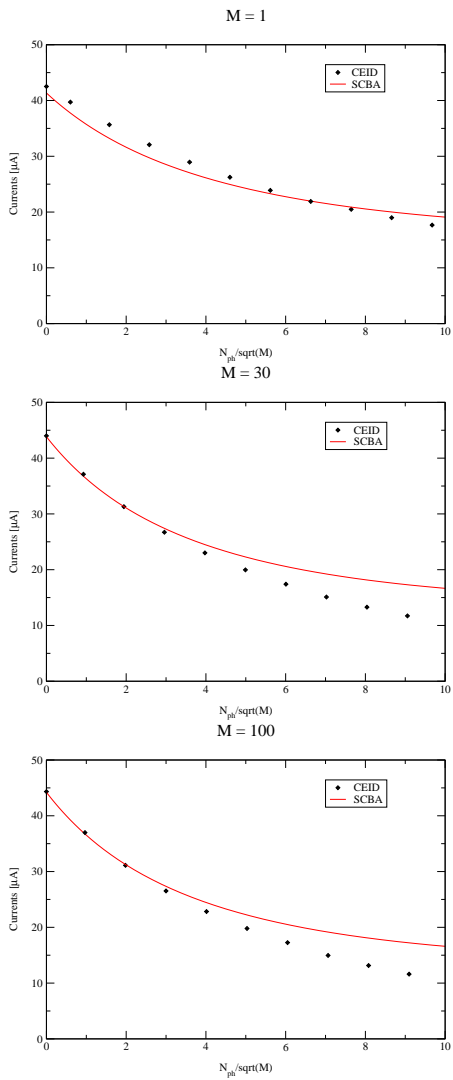

\centering
\includegraphics[totalheight = 0.2\textheight,clip=]{fig7a.eps} \\
\includegraphics[totalheight = 0.2\textheight,clip=]{fig7b.eps} \\
\includegraphics[totalheight = 0.2\textheight,clip=]{fig7c.eps}
\caption{Inelastic current as a function of $N_{\rm ph}/\sqrt{M}$ for a bias of 1V for a variety of ionic masses. (Color online)} \label{fig:current_cross}
\end{figure}

We now turn our attention to making a direct comparison of the inelastic scattering rates produced by the two methods. We consider externally damped conditions and assign the same fixed value of $C_{RR}$ to each calculation corresponding to an oscillator eigenstate with $N_{\rm ph}$ quanta. The steady-state current for a given bias (1V) is calculated as a function of $N_{\rm ph}/\sqrt{M}$ and we additionally examine how these currents vary over a range of masses. The results are shown in Fig. \ref{fig:current_cross}. We can see that the two methods remain in close agreement all the way to the point where the inelastic current has been suppressed by more than 50$\%$ relative to its value for the vibrational ground state ($N_{\rm ph} =0$). The ionic vibrational energy where more significant disagreements appear (for $N_{\rm ph}/\sqrt{M} \sim 10$) is of the order of 2 eV.

\section{Conclusions} \label{section:conclusions}
In this paper, the first direct analytical and numerical comparison has been made between the correlated electron-ion dynamics and the Self-Consistent Born Approximation. The formalisms were reviewed, and clear connections between the schemes were presented. From a numerical point of view, a model system of a linear trimer weakly coupled to two electrodes was studied, and the results are in good agreement over a range of conditions, indicating that both methods describe the underlying physics in a similar manner.

Differences begin to emerge in the limit of high thermal excitation, suggesting that, as methods for generating and summing an effective scattering series to infinite order, CEID and SCBA do ultimately differ. Furthermore, the effective Hamiltonians that the two methods use differ. The SCBA Hamiltonian is a sum of an unperturbed electronic Hamiltonian, an unperturbed phonon Hamiltonian, and a linear electron-phonon coupling. In CEID, the Hamiltonian consists of an unperturbed electronic Hamiltonian, a Hamiltonian for the bare nuclei/ions (not the phonons), and the electron-nuclear interaction. Under extreme conditions, this difference becomes exacerbated.
 
However, by applying CEID to the SCBA Hamiltonian, it was shown that CEID generates effective self-energies that, to lowest order in the electron-phonon coupling, return the Born Approximation. A challenge for further work, therefore, is to seek a general diagrammatic formulation of CEID that can be compared with NEGF to higher orders. One possible application of a diagrammatic expansion of CEID are statically disordered media, in which the effects of multiple coherent scattering are significant.  At a practical level, the agreement found between the two methods opens up the exciting possibility of combining the first-principles electron-phonon Hamiltonians, developed for use in the SCBA, with the CEID equations of motion, in order to generate corresponding dynamical electron-phonon simulations for molecular systems.

The observed behaviour of the resonant system studied here is limited by the absence of electron-electron correlation in the present model calculations. Nevertheless, on the understanding that electron-electron interactions (as well as vibrational coupling to the electrodes) may modify this behaviour, the results give the tentative indication that if, under high enough bias, the voltage window engulfs an electronic resonance, with the quasi Fermi levels of the electrodes lying in regions of low DOS, then enhanced phonon relaxation times and local heating in the resonant structure may occur, with a resultant loss of mechanical stability.

\begin{acknowledgements}
EJM, DD and TNT gratefully acknowledge funding from EPSRC for this work under grant No. EP/C006739/01. This work made use of the facilities of HPCx, the UK's national high-performance computing service, provided by EPCC at the University of Edinburgh and by CCLRC Daresbury Laboratory, and funded by the Office of Science and Technology through EPSRC's High End Computing Programme. The authors acknowledge the contribution of Cristi\'{a}n G. S\'{a}nchez, the original author of {\sc DINAMO}. TF acknowledges stimulating discussions with Andr\'es Arnau and Daniel S\'anchez-Portal, and support from FNU through grant No. 272-07-0114.
\end{acknowledgements}

\appendix

\section{Born Approximation self-energies} \label{app:baself}
In this appendix we derive the phonon contribution to the electron self-energies within the first Born Approximation, and show that they are equivalent to those obtained by substituting a current-carrying steady-state density matrix into the CEID equations.  We consider here only a single ionic degree of freedom, with mass $M$,  vibrational frequency $\Omega_0$, and electron-phonon coupling matrix
\begin{equation}
\hat{M} = \sqrt{\frac{\hbar}{2 M \Omega_0}} \frac{\partial \hat{H}_{\rm e} (R_0)}{\partial R_0} = - \sqrt{\frac{\hbar}{2 M \Omega_0}} \hat{F} .
\end{equation}
The phonon-free electronic Green's functions are 
\begin{eqnarray}
\hat{G}_0^+ (E)& = &\sum_{\alpha} \frac{| \alpha \rangle \langle \alpha |}{E - E_{\alpha} + \eye \epsilon} \\
\hat{G}_0^< (E)& = & 2 \pi \eye \sum_{\alpha}  | \alpha \rangle f_{\alpha} \delta ( E - E_{\alpha} ) \langle \alpha |.
\end{eqnarray}
Hence, from Eq (\ref{eq:ba_selfenergies1}),
\begin{eqnarray*}
\hat{\Sigma}^<_{\rm ph} (E) &=& 2 \pi \eye  \sum_{\alpha} \hat{M} | \alpha \rangle  \left [ (N_{\rm ph} + 1) \delta ( E+ \hbar \Omega_0 - E_{\alpha} ) \right. \nonumber \\
& & \left. + N_{\rm ph} \delta ( E- \hbar \Omega_0 - E_{\alpha}) \right ] f_{\alpha} \langle \alpha | \hat{M}
\end{eqnarray*}
which is the same as (\ref{eq:phonon_self_energies}).

By combining the first and third term in (\ref{eq:ba_selfenergies2}), and by performing the contour integration, one obtains
\begin{eqnarray*}
&&\frac{\eye}{2 \pi} \int ( D_0^< (\omega) + D_0^+ (\omega) ) \hat{G}_0^+ (E - \hbar \omega) \diff \omega =\\
&&  \sum_{\alpha}\left (  \frac{ (N_{\rm ph} + 1) | \alpha \rangle \langle \alpha |}{ E - E_{\alpha} - \hbar \Omega_0 + \eye \epsilon} + \frac{ N_{\rm ph} | \alpha \rangle \langle \alpha |}{E - E_{\alpha} + \hbar \Omega_0 + \eye \epsilon} \right ).
\end{eqnarray*}
The second term in (\ref{eq:ba_selfenergies2}) gives
\begin{eqnarray*}
&&\frac{\eye}{2 \pi} \int D_0^+ (\omega) \hat{G}^<_0 (E - \hbar \omega) \diff \omega =\\
& &  - \sum_{\alpha} f_{\alpha} | \alpha \rangle \left ( \frac{1}{ E - E_{\alpha} - \hbar \Omega_0 + \eye \epsilon} \right. \\
& & \left.-  \frac{1}{E - E_{\alpha} + \hbar \Omega_0 + \eye \epsilon} \right ) \langle \alpha |.
\end{eqnarray*}
By combining these terms, we obtain 
\begin{eqnarray}
\hat{\Sigma}^{\pm}_{\rm ph} (E) & = & \sum_{\alpha} \hat{M} | \alpha \rangle \left ( \frac{N_{\rm ph} + 1 - f_{\alpha}}{E - E_{\alpha} - \hbar \Omega_0 \pm \eye \epsilon} \right. \nonumber \\
& & \left. + \frac{N_{\rm ph} + f_{\alpha}}{E - E_{\alpha} + \hbar \Omega_0 \pm \eye \epsilon} \right ) \langle \alpha | \hat{M}
\end{eqnarray}
which is the same as  Eqs. (\ref{eq:phonon_self_energies}). 

\section{CEID Equations of Motion} \label{app:ceidequations}
The one-electron CEID equations of motion for non-interacting electrons are:
\begin{eqnarray}
& & \dot{\bar{R}}_{\nu} = \frac{ \bar{P}_{\nu}}{M_{\nu}} \ \ \dot{\bar{P}}_{\nu} = \bar{F}_{\nu} \label{eq:rbarpbar} \\
& & \bar{F}_{\nu} = F_{\nu}^I + \Tr \{ \hat{\rho}_{\rm e} \hat{F}_{\nu} \} - \sum_{\nu'}\Tr \{ \hat{K}_{\nu, \nu'} \hat{\mu}_{\nu'} \} \label{eq:fbar} \\
& & \dot{\hat{\rho_{\rm e}}}=\frac{1}{\eye\hbar}\,[\hat{H}_{\rm e},\hat{\rho}_{\rm e}]
-\frac{1}{\eye\hbar} \sum_{\nu} [\hat{F}_{\nu},\hat{\mu}_{\nu}]  \nonumber \\
& & +\frac{1}{2\eye\hbar} \sum_{\nu \nu'} C_{RR}^{\nu \nu'} [\hat{K}_{\nu \nu'},\hat{\rho_{\rm e}}] \label{rho} \\
& & \dot{\hat{\mu}}_{\nu} = \frac{1}{\eye \hbar} [ \hat{H}_{\rm e}, \hat{\mu}_{\nu}] + \frac{\hat{\lambda}_{\nu}}{M_{\nu}} - \frac{1}{\eye \hbar} \sum_{\nu'} C_{RR}^{\nu \nu'} [ \hat{F}_{\nu'}, \hat{\rho}_{\rm e}] \label{mu} \\
& &   {\dot {\hat \lambda}}_{\nu} = \frac{1}{\eye\hbar} [\hat{H}_{\rm e},\hat{\lambda}_{\nu}] -\frac{1}{\eye\hbar} \sum_{\nu'} C_{PR}^{\nu \nu'} [\hat{F}_{\nu'},\hat{\rho}_{\rm e}] +\frac{1}{2} ( \hat{F}_{\nu} \hat{\rho}_{\rm e} + \hat{\rho}_{\rm e} \hat{F}_{\nu}) \nonumber \\
& &- \hat{\rho}_{\rm e} \hat{F}_{\nu} \hat{\rho}_{\rm e} + \sum_{\nu' \nu''} D_{RR}^{\nu' \nu''} (\hat{\mu}_{\nu''}  {\rm Tr}\{ \hat{F}_{\nu} \hat{\mu}_{\nu'} \} - \hat{\mu}_{\nu''} \hat{F}_{\nu} \hat{\mu}_{\nu'} )\nonumber \\
& &  - \sum_{\nu'} \bar{K}_{\nu \nu'} \hat{\mu}_{\nu} -\frac{1}{2} \sum_{\nu'} (\hat{K}_{\nu \nu'} \hat{\mu}_{\nu'} +\hat{\mu}_{\nu'} \hat{K}_{\nu \nu'}) \nonumber \\
& & +\sum_{\nu'} (\hat{\mu}_{\nu'} \hat{K}_{\nu \nu'} \hat{\rho}_{\rm e} + \hat{\rho}_{\rm e} \hat{K}_{\nu \nu'} \hat{\mu}_{\nu'}) \label{lambda}\\ 
& & \dot{C}_{RR}^{\nu \nu'} = \frac{C_{PR}^{\nu \nu'}}{M_{\nu}} + \frac{C_{PR}^{\nu' \nu}}{M_{\nu'}} \label{eq:crr_eom}\\
& & \dot{C}_{PR}^{\nu \nu'} = \frac{C_{PP}^{\nu \nu'}}{M_{\nu'}} + \Tr \{ \hat{F}_{\nu} \hat{\mu}_{\nu'} \} - \sum_{\nu''} \bar{K}_{\nu \nu''} C_{RR}^{\nu'' \nu'} \label{eq:crp_eom} \\
& & \dot{C}_{PP}^{\nu \nu'} = \Tr \{ \hat{F}_{\nu} \hat{\lambda}_{\nu'} + \hat{\lambda}_{\nu} \hat{F}_{\nu'} \} \nonumber \\
& &- \sum_{\nu''} ( C_{PR}^{\nu \nu''} \bar{K}_{\nu'' \nu'} + \bar{K}_{\nu'' \nu} C_{PR}^{\nu' \nu''} ) \label{eq:cpp_eom}.
\end{eqnarray}
Above, $\bar{R}_{\nu}, \bar{P}_{\nu}$ are the mean position and momentum of the $\nu$th ionic degree of freedom, of mass $M_{\nu}$ and 
\begin{eqnarray*}
& &\hat{F}_{\nu} = - \frac{\partial \hat{H}_{\rm e} (\bar{R})}{\partial \bar{R}_{\nu}} \quad \hat{K}_{\nu \nu'} = \frac{\partial^2 \hat{H}_{\rm e} (\bar{R})}{\partial \bar{R}_{\nu} \partial \bar{R}_{\nu'}} \\
& & F_{\nu}^{\rm I} = - \frac{\partial H_{\rm I} (\hat{P}, \bar{R})}{\partial \bar{R}_{\nu}} \quad K_{\nu \nu'}^{\rm I} = \frac{\partial H_{\rm I} (\hat{P},\bar{R})}{\partial \bar{R}_{\nu} \partial \bar{R}_{\nu'}}.
\end{eqnarray*}

The second-order ionic variables are 
\begin{eqnarray*}
& & C_{RR}^{\nu \nu'} = \Tr_{\rm e} \Tr_{\rm I} \{ ( \Delta \hat{R}_{\nu} \Delta \hat{R}_{\nu'}) \hat{\rho}_{\rm eI} \} \\
& & C_{PR}^{\nu \nu'}  = \Tr_{\rm e} \Tr_{\rm I} \frac{1}{2} \{ ( \Delta \hat{P}_{\nu} \Delta \hat{R}_{\nu'} + \Delta \hat{R}_{\nu'} \Delta \hat{P}_{\nu}) \hat{\rho}_{\rm eI}\} \\
& & C_{PP}^{\nu \nu'}  = \Tr_{\rm e} \Tr_{\rm I} \{ ( \Delta \hat{P}_{\nu} \Delta \hat{P}_{\nu'}) \hat{\rho}_{\rm eI} \}.
\end{eqnarray*}
$D_{RR}$ is defined as the inverse of $C_{RR}$, such that
\begin{displaymath}
\sum_{\nu''} D_{RR}^{\nu \nu''} C_{RR}^{\nu'' \nu'} = \delta_{\nu \nu'}.
\end{displaymath}
Finally, ${\bar K}_{\nu \nu'} = K_{\nu \nu'}^{\rm I} + {\rm Tr}_{\rm  e} \{ \hat{K}_{\nu \nu'} \hat{\rho}_{\rm e} \}$.

These equations, along with the OB formalism described in Section \ref{section:ceid_plus_ob}, are  those used in Section \ref{section:applications}.

\section{Heating within the Fermi Golden Rule} \label{appendix:fgr}
As mentioned in the main text, the resonant system considered here exhibits enhanced heating under bias, with correspondingly large phonon relaxation times. In this section, we examine these phenomena qualitatively within first-order prturbation theory. Within the Fermi Golden Rule, if the electron-phonon interaction is considered as the perturbation, one can estimate the rate of energy transfer (the power) $\dot{U}$ injected into a single vibrational mode of angular frequency $\Omega_0$ \cite{montgomery2003b}:
\begin{widetext}
\begin{eqnarray}
\dot{U} & = & \frac{2 \pi \hbar ( N_{\rm ph} + 1)}{M} \sum_{\alpha, \beta = L,R} \int \diff E f_{\alpha} (E) [ 1 - f_{\beta} (E - \hbar \Omega_0)] \Tr [ \hat{D}_{\alpha} (E) \hat{F} \hat{D}_{\beta} (E - \hbar \Omega_0) \hat{F}] \nonumber \\
& & - \frac{2 \pi \hbar N_{\rm ph}}{M} \sum_{\alpha, \beta = L,R} \int \diff E f_{\alpha} (E) [ 1 - f_{\beta} ( E+ \hbar \Omega_0)] \Tr [ \hat{D}_{\alpha} (E) \hat{F} \hat{D}_{\beta} (E + \hbar \Omega_0) \hat{F} ] \label{eq:fgr_power}
\end{eqnarray}
\end{widetext}
where $M$ is the mass of the ionic degree of freedom, and indices $\alpha, \beta$ label the Lippmann-Schwinger scattering wavefunctions, originating from the respective electrodes, with occupancies $f_{\alpha}(E)$ in the Landauer picture. $\hat{D}_{\alpha} (E)$ is the partial density of states operator for the respective class of states. $\hat{F}$ is the electron-phonon coupling operator discussed earlier. Defining $U = N_{\rm ph} \hbar \Omega_0$, Eq. (\ref{eq:fgr_power}) can be rewritten as 
\begin{equation}
\dot{U}(t) = -\kappa U (t) + w_0.
\end{equation}
For the FGR calculations of maximal heating in Fig \ref{fig:heating}, the quantities $\kappa$ and $w_0$ were computed by full energy integration, from Eq. (\ref{eq:fgr_power}), with the maximal heating being given by the zero power condition $U_{\rm max} = w_0/\kappa$. 

For the purposes of gaining physical insight into the behaviour of the resonant calculation, let us now simplify the calculation as follows: we assume zero electronic temperature, and assume that the variations in the electronic Green's functions over energies of the order of $\hbar \Omega_0$ are small such that
\begin{eqnarray*}
& & \Tr \{ \hat{D}_L (E) \hat{F} \hat{D}_R (E \pm \hbar \Omega_0) \hat{F} \} \nonumber \\
& &  \approx  \Tr \{ \hat{D}_L ( E \pm \hbar \Omega_0/2 ) \hat{F} \hat{D}_R ( E \pm \hbar \Omega_0/2 ) \hat{F} \} \\
& & \pm \frac{\hbar \Omega_0}{2} \Tr \{ \hat{D}_L ( E \pm \hbar \Omega_0/2) \hat{F} \hat{D}_R' ( E \pm \hbar \Omega_0/2) \hat{F} \} \nonumber \\
& & \mp \frac{\hbar \Omega_0}{2} \Tr \{ \hat{D}_L' (E \pm \hbar \Omega_0/2) \hat{F} \hat{D}_R ( E \pm \hbar \Omega_0/2) \hat{F} \} + \mathcal{O} ( \hbar \Omega_0)^2.
\end{eqnarray*}
Hence, for $\mu_L - \mu_R \geq \hbar \Omega_0$,
\begin{eqnarray*}
\kappa &\approx& \frac{2 \pi \hbar}{M} \left [ T_{LL} (\mu_L) + T_{RR} (\mu_R) + T_{LR} (\mu_L) + T_{LR} (\mu_R) \right ] \\
& & + \frac{2 \pi \hbar}{M} \int_{\mu_R}^{\mu_L} [ \Tr \{ \hat{D}_L (E) \hat{F} \hat{D}_R' (E) \hat{F} \} \\
& & - \Tr \{ \hat{D}_L' (E) \hat{F} \hat{D}_R (E) \hat{F} \} ] \diff E + \mathcal{O}  ( \hbar \Omega_0)^2
\end{eqnarray*}
where $T_{\alpha \beta} (E) = \Tr [ \hat{D}_{\alpha} (E) \hat{F} \hat{D}_{\beta} (E) \hat{F} ]$ and $\mu_{L,R}$ are the electrochemical potentials of the left and right battery terminals. For a system with reflection symmetry about the origin (which we can assume here), the terms in the integral cancel identically, and hence $\kappa$ and $w_0$ are given by
\begin{eqnarray}
\kappa &=& \frac{2 \pi \hbar}{M} \left [  T_{LL} ( \mu_L) + T_{RR} (\mu_R) +  T_{LR} (\mu_L) + T_{LR} ( \mu_R)  \right ]   \nonumber \\
w_0 &=& \frac{2 \pi \hbar}{M} \int_{\mu_R + \hbar \Omega_0/2}^{\mu_L-\hbar \Omega_0/2} T_{LR} (E) \, \diff E
\end{eqnarray}
Consider now a situation in which under a large enough bias, the energy window for conduction engulfs an electronic resonance, with $\mu_L$ and $\mu_R$ now lying in regions of low densities of states, on each side of the resonance. Then $\kappa$, which collects contributions from energies in  the vicinity of the two Fermi levels, is small, and gets smaller with increasing bias, as $\mu_L$ and $\mu_R$ move further away from the resonance. Since $\kappa$ depends quadratically on the density of states, its decrease with bias should be faster than linear. $w_0$, on the other hand, collects contributions from the entire conduction window, and saturates with increasing bias. This is the origin of the enhancement of $U_{\rm max}$ and of the phonon equilibration time $\tau = \kappa^{-1}$ compared with the ballistic case (in which $w_0$ increases linearly with bias, and $\kappa$ is approximately bias-independent \cite{mceniry2007a}).


\bibliography{080228-arxiv.bib}
\end{document}